\documentclass[12pt]{article}
\usepackage{amssymb,amsmath,amsfonts}
\begin{document}

\begin{titlepage}

                            \begin{center}
                            \vspace*{2cm}
\large\bf{Distributivity and deformation of the reals from Tsallis entropy}\\

                            \vfill

              \normalsize\sf    NIKOS \  KALOGEROPOULOS$^\ast$\\

                            \vspace{0.2cm}
                            
 \normalsize\sf Weill Cornell Medical College in Qatar\\
 Education City,  P.O.  Box 24144\\
 Doha, Qatar\\

                            \end{center}

                            \vfill

                     \centerline{\normalsize\bf Abstract}
                     
                           \vspace{3mm}
                     
\normalsize\rm\setlength{\baselineskip}{18pt} 

\noindent We propose
a one-parameter family \ $\mathbb{R}_q$ \ of deformations  of the reals,  which is 
motivated by  the generalized additivity of the Tsallis entropy. We introduce a generalized 
multiplication which is distributive with respect to the generalized addition of the Tsallis entropy.
These operations establish a one-parameter family of field isomorphisms  \ $\tau_q$ \ 
between \ $\mathbb{R}$ \ and \ $\mathbb{R}_q$ \  through which  an absolute value
on \ $\mathbb{R}_q$ \ is introduced. This  turns out to be a  quasisymmetric map, 
whose metric and measure-theoretical implications  are pointed out.\\    

                             \vfill

\noindent\sf  PACS: \  \  \  \  \  \  02.10.Hh, \  05.45.Df, \  64.60.al  \\
\noindent\sf Keywords: \ Tsallis entropy, \ Nonextensive entropy, Nonextensive Statistical Mechanics.   \\
                             
                             \vfill

\noindent\rule{8cm}{0.2mm}\\
\noindent \small\rm $^\ast$  E-mail: \ \  \small\rm nik2011@qatar-med.cornell.edu\\

\end{titlepage}

%%%%%%%%%%%%%%%%%%%%%%%%%%%%%%%%%%%%%%%%%%%%%%%%%%%%%%%%%%%%%%%%%%%%

                            \newpage

\normalsize\rm\setlength{\baselineskip}{18pt}

          \centerline{\large\sc 1. \ Introduction}

                                     \vspace{3mm}

    The Tsallis entropy [1], [2] is a thermodynamic potential
aiming to provide an alternate entropic form to that of Boltzmann/Gibbs/Shannon (BGS) on which the equilibrium Statistical 
Mechanics relies. If one assumes a system with a discrete set of states indexed by \ $i$, \ occurring with 
corresponding probabilities \ $p_i$, \  the Tsallis entropy, is defined by
\begin{equation}
    S_q =  k \ \frac{1}{q-1} \left( 1- \sum_i p_i^q \right)
\end{equation}  
where \ $k$ \ stands for Boltzmann's constant. In case the system has a continuous set of states 
described by a probability distribution \ $p(x)$, \ where \ $x$ \ takes values in an appropriate parameter space \ $M$, \ then the expression
for the Tsallis entropy becomes  
\begin{equation}
   S_q = k \  \frac{1}{q-1}  \left(1 - \int _M [p(x)]^q \ dx\right)
\end{equation}
We will assume in the sequel, for simplicity, that \ $k=1$. \ The Tsallis entropy describes the thermodynamic behavior of 
systems that are either out equilibrium, or  exhibiting spatial or temporal long-range correlations, or 
whose phase space portraits are fractals ([2] and references therein). This list is not all-inclusive though. In the cases mentioned, 
and despite the striking successes of the  BGS entropy, there is no reason why the BGS entropy should correctly describe the 
thermodynamic behavior of such systems.   
On the other hand, there are very diverse phenomena among the cases mentioned above whose collective behavior may be described by the 
Tsallis entropy. This is indicated by data fits as well as  analytical arguments and numerical results [2]. What is lacking though is the concrete 
dynamical basis for the Tsallis entropy similar to the molecular chaos hypothesis (stosszahlansatz) in kinetic theory or to the ergodic hypothesis 
of the BGS case. This lack of understanding of Tsallis's entropy dynamical foundations, certainly does not invalidate its use or range of applicability. 
We may recall, by comparison, that the BGS 
entropy has  not really been rigorously derived for cases of physical interest, despite its striking predictions throughout its existence for 
almost a century and a half. A related problem is that of the, a priori,  determination the value of the non-extensive parameter \ $q$ \ for Hamiltonian 
or dissipative systems. Some results in this direction exist [2], but they are sporadic, the overall picture being still unclear.
The fact that each system described by the Tsallis entropy seems to possess several (at least three, if not infinite [2]) values for its $q$ parameter,
depending on which aspect of it is studied, complicates further this task. \\

 In the present work, we address one particular issue pertaining to the Tsallis entropy, which may eventually help gain a better understanding 
of the meaning and ways to calculate the value of  \ $q$: \ that of  creating an algebraic structure that 
reflects the composition properties of the Tsallis entropy and the systems that it purports to describe. In Section 2, we construct a 
one-parameter family of deformations of the reals and define a generalized multiplication that is distributive with respect to the addition 
induced by the composition of the Tsallis entropies. In Section 3, we point out that this family of maps is quasisymmetric, and comment 
on its relation to fractals, which motivated the construction of the Tsallis entropy. Section 4 contains  further remarks regarding the 
generalization and possible implications of these deformation maps in higher dimensions, to be pursued in future work.\\            

After the present work was completed, we became aware of the already published paper [3] which had constructed, clearly preceeding us, 
the same algebraic structure as the present work. We believe that the algebraic construction of the present work is the same as that of [3], 
but could not prove explicitly this  conjectured  equivalence. This conjectured equivalence amounts to finding an explicit isomorphism, if it exists, 
between the two generalized multiplications, eq. (24) of [3] and eq. (33) of the present work, something that this author was unable 
to readily accomplish.\\  

                                  \vspace{5mm}

%%%%%%%%%%%%%%%%%%%%%%%%%%%%%%%%%%%%%%%%%%%%%%%%%%%%%%%%%%%%%%%%%%%

          \centerline{\large\sc 2. \ Definition and properties of \  $\mathbb{R}_q$ }

                                        \vspace{3mm}

 One can immediately check that the Tsallis entropy obeys the composition law
\begin{equation} 
 S_q (A + B) = S_q(A) + S_q(B) + (1-q) S_q(A)S_q(B)
\end{equation}
where systems \ $A$ \ and \ $B$ \ are assumed to be independent, in the conventional sense of the word, namely
their probability distributions \ $p^A$ \ and \ $p^B$ \ respectively, obey the relation   
\ $p^{A+B} = p^A p^B$. \ It may be instructive to contrast this with the BGS
expression for entropy which is additive [4], namely it obeys  \ $S(A+B) = S(A) + S(B)$ \ for the two systems 
\ $A$ \ and \ $B$ \ with probability densities obeying the same independence relation as above.   \\

 Motivated by (3), one can reconsider the rule for the composition of 
systems described by the Tsallis entropy, and thus define a generalized addition  by
\begin{equation}
    x \oplus _q  y = x + y + (1-q) xy
\end{equation} 
for any \ $x, y \in \mathbb{R}$ \ and any \ $q\in\mathbb{R}$, \  where \ $\mathbb{R}$ \  denotes 
the set of real numbers. This generalized addition has, as also expected from its name,  the 
usual algebraic properties: associativity, commutativity, zero and opposite elements,  
under which \ $(\mathbb{R}, \oplus_q) $ \  becomes an Abelian group. On the other hand, the 
definition of a corresponding multiplication is somewhat less obvious. One such definition 
[5], [6]
\begin{equation}
  x \otimes_q y = (x^{1-q} + y^{1-q})^\frac{1}{1-q}
\end{equation}
was motivated by the generalized additivity property of the $q$-logarithm. This was motivated by the desire to 
make the Tsallis entropy look as much as possible like the BGS entropy, and is defined as
\begin{equation}
 \ln_q x = \frac{x^{1-q} - 1}{1 - q}, \ \  \  x \in \mathbb{R}_+
\end{equation}
The $q$-logarithm can be seen to obey the additivity property
\begin{equation}
 \ln_q (x \otimes_q y) = \ln_q x + \ln_q y
\end{equation}     
Unfortunately, \ $\otimes_q$ \ does not obey the usual distributive property with respect to \ $\oplus_q$. \
Although there is no physical reason why this may be considered a drawback, it is worth recalling that
most sets possessing two operations that have been useful in Physics have, at least, a basic structure of rings, 
fields or vector spaces, if no more structure is present. Hence it would be quite desirable to construct a generalized 
multiplication that is distributive with respect to the generalized addition. Such an attempt was made in [7] by using the 
generalized multiplication provided above as generalized addition, and introducing  an essentially 
new generalized multiplication. That approach however, and the ensuing formal definitions of a derivative 
and an integral,  did not recover the usual operations in the limit \  $ q \rightarrow 1 $ \ and this may be one reason 
why they did not gain any further traction.\\ 

The question of whether one can still define a generalized addition and multiplication that obey, in some sense, 
the distributive property had remained open for a while [8], but a recent proposal [3] has provided a satisfactory answer. 
If one maintains the form of the generalized multiplication 
given above, and attempts to modify the generalized addition, an argument exists that satisfying the distributive property 
may not even be possible [2]. On the other hand, an algebraic structure requiring the validity of the distributive property, 
akin to a ring, field or vector space is highly desirable [8]. This a basic motivation guiding the algebraic construction of [3] 
and of the present work. \\

In [3], the algebraic construction of Section 3 is preceded by several quite interesting and pertinent comments of, mostly, set theoretical 
and arithmetic nature, having no counterpart in  our work.  
Proceeding to the algebraic construction, the authors of [3] kept the 
generalized addition (4) intact, 
as we do in the present work. To enforce the validity of the  distributivity property, they started by relying on (4) and deformed the 
underlying set, as we also do in the present work. 
They then searched for a generalized multiplication/product  \ $\Diamond_q$ \  in the set they constructed, 
which we call \  $\mathbb{R}_q$ \  in the present work,  which  is a group homomorphism. 
More explicitly, eq. (6) of [3] demanded that 
\begin{equation}   
        x_q \Diamond_q  y_q = (xy)_q
\end{equation}
This property was fulfilled by defining, in eq. (24) of [3], the generalized multiplication as 
\begin{equation}
    x \Diamond_q y = \frac{(2-q)^{\frac{\log [1+(1-q)x ] \log [1+(1-q)y ] }{ [ \log (2-q) ]^2}} -1}{1-q}
\end{equation}
which was subsequently checked to satisfy the group homomorphism property (8), reduce to the usual multiplication 
for \ $q\rightarrow 1$, \  and indeed obey the distributivity property with respect to addition, namely 
\begin{equation}
 x_q \Diamond_q (y_q \oplus_q z_q) = (x_q \Diamond_q y_q) \oplus_q (x_q \Diamond_q z_q)
\end{equation} 
Beyond this point, [3] took a completely different path from the one we will follow in the sequel.\\

Our approach follows a route that is very similar to that of [3] overall, as far as the algebraic constructions are concerned. 
The only essential difference at the algebraic level, is between the generalized multiplication (33)   
that we propose in this paper and the generalized product (9) which was introduced in [3]. We think that the form 
of the generalized  addition \  $\oplus_q$ \  is quite fundamental, as it emerges from the composition of the Tsallis entropy, 
hence, as was also done in [3],  it should be kept intact. As a result, and in complete accordance with the results of [3],  we deform the underlying set 
\  $\mathbb{R}$ \ in such a way as to maintain the validity of \  $\oplus_q$. \  We then construct the deformation \ $\mathbb{R}_q$ \ in such a way that the  
distributive property is already built-in. So, we  
construct, in several steps, a one-parameter family of deformation maps \ $\tau_q : \mathbb{R} \rightarrow \mathbb{R}_q$ \ as follows. We assume that
\ $\tau_q: 0 \rightarrow 0_q$ \ and that \ $\tau_q: 1 \rightarrow 1_q$. \  We then continue building the elements that comprise \ $\mathbb{N}_q$ \
by using \ $\oplus_q$, \ instead of \ +, \ as is done in elementary arithmetic. We arrive at  the recursion relation 
\begin{equation}   
   \tau_q(n) \equiv n_q = (2-q)n_{q-1} +1, \ \ \ \ 1_q = 1
\end{equation}
Since (11) is first-order inhomogeneous, 
we transform it into a second-order homogeneous recursion relation by subtracting by parts two consecutive terms. As a result, we get
\begin{equation}
   n_q = (3-q) (n-1)_q + (2-q) (n-2)_q, \ \ \ \ n \geq 3, \ \ \ \ 1_q = 1, \ \  2_q = 3-q 
\end{equation}
Its characteristic polynomial is
\begin{equation} 
   x^2 - (3-q)x + (2-q) = 0
\end{equation}
therefore its general solution has the form
\begin{equation}
    n_q =  c_1 \left( \frac{3-q + \sqrt{4 + (2-q)^2}}{2}\right)^n + c_2 \left( \frac{3-q - \sqrt{4 + (2-q)^2}}{2}\right)^n 
\end{equation}
To determine the constants \ $c_1$ \ and \ $c_2$, \ we substitute the general form into the expressions for \ $1_q$ \ 
and \ $2_q$, \ and we eventually find
\begin{equation} 
   n_q = \frac{1}{\sqrt{4 + (2-q)^2}} \left[ \left( \frac{3-q + \sqrt{4 + (2-q)^2}}{2}\right)^n +  \left( \frac{3-q - \sqrt{4 + (2-q)^2}}{2}\right)^n \right]
\end{equation}
for \ $n \geq 3$. \  We proceed by determining the ``opposite" values of \ $n_q$, \ which are indicated by 
$\ominus_q n_q \equiv (\ominus n)_q$, \ by demanding that 
\begin{equation}
  n_q \oplus_q (\ominus_q n_q) = 0_q
\end{equation}
which gives  
\begin{equation}
   (\ominus n)_q = - \frac{n_q}{1+(1-q)n_q}
\end{equation}
which can, inductively, be seen to imply that 
\begin{equation}
  (\ominus n)_q = - \frac{n_q}{(2-q)^n}
\end{equation}
Since, by construction
\begin{equation}
 n_q \oplus_q (\ominus n)_q = 0_q
\end{equation}
by using (4), \ we get
\begin{equation}
   (2-q)^n - 1 - (1-q)n_q = 0
\end{equation}
which implies 
\begin{equation}
    n_q = \frac{(2-q)^n - 1}{1-q}, \ \ \ \  n \in \mathbb{N} 
\end{equation} 
So,  as a by-product of the process of construction of \ $\mathbb{R}_q$, \ we have found the Diophantine-like expression
\begin{equation}
\frac{1}{\sqrt{4 + (2-q)^2}} \left[ \left( \frac{3-q + \sqrt{4 + (2-q)^2}}{2}\right)^n +  \left( \frac{3-q - \sqrt{4 + (2-q)^2}}{2}\right)^n \right]
   =  \frac{(2-q)^n - 1}{1-q}
\end{equation}
 We can readily verify that 
\begin{equation}
   \lim_{q\rightarrow 1} \ n_q = n 
\end{equation}
and 
\begin{equation}
   (n+m)_q = n_q \oplus_q m_q 
\end{equation}
which implies  
\begin{equation}
   \lim_{q\rightarrow 1}  \  (n+m)_q = n+m
\end{equation}
So, we can extend \ $\mathbb{N}_q$ \ to \ $\mathbb{Z}_q$, \ which becomes an Abelian 
group under \ $\oplus_q$. \ We also notice, in passing,  that 
\begin{equation}
   n_q \ominus_q m_q = \ominus_q (m_q \ominus_q n_q)
\end{equation}
and that \ $\ominus_q$ \ is idempotent, namely that 
\begin{equation}
       \ominus_q (\ominus_q n_q) = n_q,  \ \ \ \forall \  n_q \in \mathbb{Z}_q
\end{equation}
The non-trivial step is the definition of a generalized multiplication \ $\otimes_q$ \ in \ $\mathbb{Z}_q$. \ 
We demand that the map \ $\tau_q$ \ be a homomorphism with respect to multiplication, namely 
\begin{equation}
     \tau_q(n) \otimes_q \tau_q(m) = \tau_q (nm)
\end{equation}
or in our shorthand notation 
\begin{equation}
     n_q \otimes_q m_q = (nm)_q, \ \ \ \ \forall \ n, m \in \mathbb{Z}
\end{equation}
This homomorphism guarantees that the distributivity property will automatically hold in \ $\mathbb{Z}_q$, \ as it will be the image 
under \ $\tau_q$ \ of the usual distributivity property in \ $\mathbb{Z}$. \  To determine such a multiplication we start by observing that, 
upon a binomial expansion,
\begin{equation}
    n_q = \sum_{k=0}^{n-1}  {n \choose k} (1-q)^{n-1-k}
\end{equation}
We consider the polynomial ring in one indeterminate \ $\mathbb{Z}[q]$ \ whose elements are the expressions of the elements
\ $n_q \in \mathbb{Z}_q$ \  as polynomials of \ $1-q$. \  This expansion is formalized by the map \ $\nu: \mathbb{Z}_q \rightarrow \mathbb{Z}[q]$ \
which is clearly invertible.  Then, we define the ``projection" map \ $\pi: \mathbb{Z}[q] \rightarrow \mathbb{Z}[q]$ \ 
of this polynomial to its highest power monomial, by 
\begin{equation}
  \pi  \left\{ \sum_{k=0}^{n-1}  {n \choose k} (1-q)^{n-1-k} \right\}  = (1-q)^{n-1}
\end{equation}
We also define the ``completion" map \ $\sigma: \mathbb{Z}[q] \rightarrow \mathbb{Z}[q]$ \  by
\begin{equation}
     \sigma \left\{ (1-q)^{n-1} \right\}  = \sum_{k=0}^{n-1}  {n \choose k} (1-q)^{n-1-k}
\end{equation}
The generalized multiplication  \ $\otimes_q$ \ is then defined by
\begin{equation}
   n_q \otimes_q m_q = \nu^{-1} \sigma \left\{ \left[ (\pi\nu n_q)^\frac{1}{n-1} (\pi\nu m_q)^\frac{1}{m-1}\right]^\frac{nm-1}{2} \right\} 
\end{equation}
With this definition we can easily check that (28)/(29) is indeed satisfied. This automatically implies that \ $\tau_q$ \ is a one-parameter family 
of ring homomorphisms between \ $(\mathbb{Z}, +, \cdot )$ \ and \ $(\mathbb{Z}_q, \oplus_q, \otimes_q) $ \ since \ $\otimes_q$ \ is easily checked to be 
associative,  commutative and have \ $1_q$ \ as unit. Moreover the \ $\otimes_q$ \ is trivially distributive with respect to \ $\oplus_q$  \
 due to (28)/(29). The characteristic of \ $(\mathbb{Z}_q, \oplus_q, \otimes_q)$ \ is zero since 
\begin{equation}
     1_q \oplus_q \ldots 1_q = 0_q   
\end{equation}
amounts to \ $n_q = 0_q$ \ which is generically true only for \ $n=0$. \ These definitions allow us to define a division for all non-zero elements of 
\ $(\mathbb{Z}_q, \oplus_q, \otimes_q)$, \ henceforth to be indicated only by its defining set. Indeed, the inverse element of \ $n_q \in \mathbb{Z}_q$ \
is given by
\begin{equation}  
   (n^{-1})_q = \frac{(2-q)^\frac{1}{n} - 1}{1-q}
\end{equation} 
which can be readily verified using the definition of  \ $\otimes_q$. \ Then we can extend the definition of \ $\mathbb{Z}_q$ \ to \ $\mathbb{Q}_q$ \ 
as follows 
\begin{equation}  
  \left( \frac{n}{m} \right)_q = \frac{(2-q)^\frac{n}{m} - 1}{1-q}, \ \ \ \ \forall \ n,m \in \mathbb{Z}, \  \  \  m \neq 0
 \end{equation} 
 The following clarifying comment may be worth making at this point. The expansion of \ $n_q$ \  given in (30) does not hold for \ 
 $n \not\in \mathbb{N}$  \  as the corresponding polynomial needs to be substituted by an infinite series. Then issues of convergence, such as determination 
 of the radius of convergence or the degree of regularity of the ensuing expansion needed for some useful algebraic properties to hold, will have to be 
 settled.  Moreover the 
 multiplication rule (33) cannot be straightforwardly applied as it uses in an essential way the existence of a highest degree monomial in \ $\mathbb{Z}_q$. \    
 To sidestep such difficulties we utilize the map \ $\tau_q$. \ Through it, the multiplication of any two numbers, not only belonging in \ $\mathbb{Q}_q$ \
 but also in \ $\mathbb{R}_q$, \ which is the (Dedekind) closure of the former set, can be defined by passing to the corresponding multiplication of elements 
 of \ $\mathbb{Q}$ \ or \ $\mathbb{R}$. \  Then we can easily extend  \ $\tau_q: \mathbb{Z} \rightarrow \mathbb{Z}_q$ \  to \ $\tau_q: \mathbb{R} \rightarrow 
 \mathbb{R}_q$, \
 especially when \  $\tau_q$ \  is injective. A sufficient condition for this to occur is when \ $\tau_q$ \ is monotonic.\\
 
  We will focus from now on in the case of \ $0\leq q \leq1$ \ for the following reasons. 
 First, we are interested in \ $q\geq 0$, \ since for \ $q<0$ \ zero-probability states will 
 give rise to infinities as can be seen from (1). This may be easily remedied if we restrict the index set by choosing, beforehand, to exclude all zero probability 
 states  from the sum of 
 (1). Notice though that for \ $q<0$ \ the Tsallis entropy becomes minimum at equal probabilities, as opposed to the BGS entropy which gets maximized. This 
 may not be
 a major drawback if we remember to substitute max for min in the Tsallis case for \ $q<0$. \  In the same vain, the Tsallis entropy is convex for \ $q<0$ \ in 
 contrast to the 
 BGS entropy  which it is concave, a difference that again can be easily eliminated by a word substitution. What may be far more important though, is that the 
 definition of 
 experimental robustness (Lesche-stability) [9], [10] has been verified 
 to hold for the Tsallis entropy only for \ $q>0$ \ [11]. Second, following the spirit of the BGS expression where the entropy of non-independent 
 systems is generally super-additive \  $S(A+B) \geq S(A) + S(B)$, \  we have superadditivity for the Tsallis entropy, as can be seen from (3), 
 when \ $q\leq 1$. \  
 Moreover, if we use the constraint of given mean energy, (with the mean either computed with respect to the probability distribution [1] or to its ``escort" 
 counterpart [12], 
 [13]), and extremize the 
 Tsallis entropy, one  finds [1] that the ensuing probability distribution exhibits an asymptotic power-law behavior only when \ $q<1$, \  as opposed to the 
 existence of a cut-off
 when \ $q>1$. \  We consider the former behavior quite a bit more interesting for further examination than the latter, subjective as such a decision might be. 
 Then, for any \ $q \in [0, 1]$, \ the function 
 \  $\tau_q(x)$ \  is smooth and increasing due to \ $\frac{d\tau_q(x)}{dx} > 0$, \   therefore  monotonic, thus the corresponding one-parameter family of 
 homomorphisms can 
 indeed be consistently extended over \  $\mathbb{R}$, \  as has been tacitly assumed so far, thus justifying the comments at the end of the previous  
 paragraph. 
 In this case the maps \ $\tau_q$ \ are bijective and invertible, hence they form a one-parameter family of field isomorphisms of  \ $\mathbb{R}$. \\
 
   Following the same approach we can, in turn, define imaginary numbers and 
 eventually a one-parameter family of deformed complex numbers \ $\mathbb{C}_q$ \ which much like \ $\mathbb{C}$ \ is also algebraically closed,  being 
 the image of 
 \ $\mathbb{C}$ \ under the field isomorphism \ $\tau_q$. \ Using \ $\tau_q$ \ we can also define an order relation \ $>_q$ \ between any two elements of  \ $
 \mathbb{R}_q$ \ by
 \begin{equation}
 x_q >_q y_q \  \  \  \mathrm{iff} \ \  \ x > y        
 \end{equation}
 thus turning \ $\mathbb{R}_q$ \ into a totally ordered set. The total order \ $>_q$ \ allows the definition of \ $\mathbb{R}^\oplus _q = \left \{ x_q \in \mathbb{R}
 _q : 
 x_q >_q 0_q \right\} $ \ which, not too surprisingly, can also be characterized as \ $\mathbb{R}^\oplus _q = \tau_q \ \mathbb{R}^+$ \ Continuing in this spirit, 
 we can  
 define the  $q$-absolute value on \ $\mathbb{R}_q$ \ as the evaluation map \ $|\cdot |_q : \mathbb{R}_q \rightarrow \mathbb{R}^\oplus_q$ \ given by  
 \begin{equation} 
       | x_q \ominus_q y_q |_q = \tau_q (|x-y|) 
 \end{equation}
 or  more explicitly
 \begin{equation}
      | x_q \ominus_q  y_q |_q = \frac{(2-q)^{|x-y|} - 1}{1-q}
 \end{equation} 
 
  It may be worth mentioning at this point that one can define exponential and logarithmic functions on \ $\mathbb{R}_q$ \ which however are \emph
{different} 
 from the $q$-exponential and $q$-logarithm whose definitions were motivated by the Tsallis entropy [2]. Indeed, setting temporarily aside issues of 
 convergence, one can 
 formally define \ $ \mathrm{EXP}_q: \mathbb{R}_q \rightarrow \mathbb{R}^\oplus _q $ \ by  
 \begin{equation}
        \mathrm{EXP}_q(x_q) = \tau_q (\exp \ x)
 \end{equation}
 which can also be expressed as
 \begin{equation}
        \mathrm{EXP}_q = \frac{(2-q)^{\exp x} - 1}{1-q} 
 \end{equation}
 Since \ $\tau_q$ \ is a field isomorphism for each value of \ $q\in [0, 1]$ \ we can also define the inverse of \ $\mathrm{EXP}_q$ \ denoted by \ 
 $\mathrm{LOG}_q : \mathbb{R}^\oplus _q \rightarrow \mathbb{R}_q$ \
 by demanding that 
 \begin{equation}
     \mathrm{EXP}_q(\mathrm{LOG}_q (x_q)) = \mathrm{LOG}_q(\mathrm{EXP}_q (x_q)) = x_q
 \end{equation}
 Such a logarithmic function can be easily obtained as the image of the natural logarithmic function \ $\ln $ \ on \ $\mathbb{R}^+_q$ \ as
 \begin{equation} 
         \mathrm{LOG}_q (x_q) = \tau_q (\ln \ x) 
 \end{equation}
 or more explicitly
 \begin{equation}
         \mathrm{LOG}_q (x_q) = \frac{(2-q)^{\ln x} -1}{1-q}
 \end{equation}
 
After defining the generalized multiplication \ $\otimes_q$ \ in (33), the subsequent algebraic structure discussed above follows straightforwardly.
What we have done is to  construct isomorphic fields \ $\mathbb{R}_q$ \ which are, by definition, algebraically indistinguishable, and whose identification is 
realized through the field isomorphisms \ $\tau_q$. \  When we remind ourselves that in  Statistical Mechanics we tend to work, mostly,
with the usual sets \ $\mathbb{Z}$, \ $\mathbb{R}$, \ $\mathbb{C}$ \ we see that what have have done is to have constructed (algebraically identical) 
deformations of them, whose properties however reflect more closely the composition properties of the Tsallis entropy than of that of the BGS entropy. 
The most dramatic implications though of the deformation maps \ $\tau_q$ \ appear at the level of the metric and measure-theoretical properties
of \ $\mathbb{R}_q$ \ to which we briefly turn our attention in the next section.\\

                                                             \vspace{5mm}

%%%%%%%%%%%%%%%%%%%%%%%%%%%%%%%%%%%%%%%%%%%%%%%%%%%%%%%%%%%%%%%%%%%%%%%%%%%

\centerline{\large\sc 3. \ Metric and measure through  \  $\tau_q$ }

                                        \vspace{3mm}

  The structure of Statistical Mechanics relies in an essential way on two fundamental concepts: metric and measure. It is actually the 
latter which is, in a sense much more important. These are completely distinct
concepts, which under some restrictive conditions become inter-related. As an example, consider the Euclidean metric on \ $\mathbb{R}^n$. \ Then 
Lebesgue's theorem states that there is a unique translation invariant measure, which we tend to call ``volume". The construction can be repeated for 
Riemannian manifolds \ $M$, \ which after all, are not only locally homeomorphic, but also locally isometric (to zeroth order) to Euclidean spaces. 
Then there is a there is a canonical measure, called  ``volume" \ $\mathrm{Vol}$, \ which is uniquely characterized by the following two properties [14]: \\ 
i)  \  For surjective, distance decreasing maps \ $f: M_1 \rightarrow M_2$, \ it satisfies \ $\mathrm{Vol} \ M_1 \leq \mathrm{Vol} \ M_2$. \\
ii)  The unit cube in \ $\mathbb{R}^n$ \ is normalized to have volume equal to one.\\  
In general, however, no such nice relation between metric and measure exists. Consider, for instance, the case of  finite-dimensional Finsler (Minkowski) 
spaces, which are linear spaces endowed with a metric induced by a Banach norm. Then a variety of distinct but ``reasonable" definitions exist for 
the definition of volume [15] none of which is more ``natural" than the others from the viewpoint of the existing metric.\\

 Motivated by the above, it may be of some interest for future applications to physical systems described by the Tsallis entropy to examine the metric 
and measure-theoretical aspects of the \ $\tau_q$ \ isomorphisms. A simple, but important, observation is that the maps \ $\tau_q$ \ are not isometries, 
meaning that they do not preserve  distances (the metric structure), namely
\begin{equation}      
     \tau_q (|x-y|) =  |x_q \ominus_q y_q|_q \neq |x-y| 
\end{equation} 
If one relaxes the requirement for invariance of distances by a map, a more general class of maps of interest are those that distort the metric by a finite factor,
namely the (bi-) Lipschitz maps. A direct observation shows that \ $\tau_q$ \ are not even Lipschitz, namely that there is no constant \ $C>0$ \ such that 
\begin{equation}
      |x_q \ominus_q y_q|_q = \frac{(2-q)^{|x-y|} - 1}{1 - q} \leq C|x-y|
\end{equation}
 It is not really hard to understand the reason why the Lipschitz condition is not satisfied.
\ $\tau_q$ \ are exponential maps and one can see that the exponential function grows uncontrollably large, so it cannot 
obey the Lipschitz property. This is also obvious from the graph of the exponential function which becomes progressively steeper without having any 
upper bound.  A still more general category of maps between metric spaces is that of quasisymmetric ones [16]. We start with a slightly more general 
definition. Let \ $M$ \ and \ $N$ \ be general metric spaces with metrics indicated, for either space, using the Polish notation \ $|x-y|$. \ An embedding
\  $f: M \rightarrow N$ \ is called weakly C-quasisymmetric, if there is a constant \ $C\geq 1$ \ such that 
\begin{equation}
     |x - y|  \leq |x - z| \ \ \ \ \  \mathrm{implies} \ \ \ \ \  |f(x) - f(y)|  \leq C \ |f(x) - f(z)|, \ \ \ \ \ \ \forall \ x,y,z \ \in M
\end{equation} 
We can straightforwardly confirm that \ $\tau_q, \ \  q \in [0, 1]$, \ is a one-parameter family of weakly 1-quasisymmetric maps.  It 
turns out [17] that a weakly quasisymmetric embedding \ $f: \mathbb{R} \rightarrow \mathbb{R} $ \ such as \ $\tau_q$, \ is quasisymmetric. 
An embedding of metric spaces \ $f: M \rightarrow N$, \  as above, is called $\eta$-quasisymmetric, if there is a homeomorphism 
\ $\eta : [0, \infty ) \rightarrow [0, \infty )$ \ such that
\begin{equation} 
      \frac{ |f(x) - f(y)| }{|f(x) - f(z)|} \leq \eta \left( \frac{|x - y|}{|x - z|} \right)
\end{equation}
for any \ $x, y, z \in M$ \ and for all \ $t>0$ \ [16].  As can be seen from this definition, quasisymmetric maps can stretch distances a lot, but there is a
bound on how much they can stretch relative distances. In a sense, they tend to preserve the shape of an object, although they may considerably change 
its size.
By contrast, Lipschitz maps, are more ``tame" in that they tend not to distort too much either the shape or the size of an object. It follows from the definitions,  
that  Lipschitz maps (or, to be more precise, $C$-bi-Lipschitz maps) are \ $C^2t$-quasisymmetric.\\

  A class of measures that has a close relation to 
quasisymmetric maps are the doubling measures. They emulate, in a sense, the close relation/compatibility  
between Euclidean (and Riemannian) metrics and measures.     
Doubling measures are (Borel regular) measures \ $\mu$ \ on  general metric-measure spaces \ $X$ \ that satisfy the following condition: 
Each ball with center \ $x\in X$ \  of radius \ $2r$, \ (with \ $r$ \ being smaller than half of the diameter of \ $X$, \ if the diameter of \ $X$ \ is finite)     
can be covered by at most \ $c$ \ balls of radius \ $r$, \ or more concretely
\begin{equation}
          \mu (B_{2r}(x)) \leq c \  \mu (B_r(x))
\end{equation} 
where the constant \ $c$ \ may depend on \ $\mu$, \ but does not depend on \ $x\in X$. \ Here \ $B_r(x)$ \ indicates an open ball (open interval in the case of
 \ $ \mathbb{R}$) centered at \ $x$ \ of radius \ $r$. \ Thus the doubling condition is a condition which is used to control the behavior of measures in a uniform 
 way at all scales in \ $X$. \  A simple example of doubling measure is the volume in \ $\mathbb{R}^n$. \ A space in which the volume is not doubling is the 
 hyperbolic space \ $\mathbb{H}^n$ \ since the volume of balls increases exponentially with the radius, as the radius of these balls increases. The set \ 
 $\mathbb{R}$, (equivalently: the real line) is quite special when it comes to quasisymmetric maps. Indeed, there is an injective correspondence between 
 quasisymmetric maps \ $f$ \ on the real line and doubling measures \ $\mu$ \ on it [18], given by
 \begin{equation}
      f(x) = \int_0^x d\mu (t)
 \end{equation}   
This relation can be, conversely, interpreted as stating that if any doubling measure on the line \ $d\mu$ \ is integrated, it will result in a quasisymmetric 
homeomorphism \ $f$ \ of the line, which is 
such that the given doubling measure \ $\mu(I) $ \ will be the pullback of the Lebesgue measure \ $\mathrm{Vol}f(I), \  I\subset \mathbb{R}$. \ This statement 
can be 
generalized by proving that the quasisymmetric image of a doubling space is doubling [17]. So, the doubling condition is a quasisymmetric invariant. 
Here ``doubling space" refers to a metric space \ $M$ \ with the 
property: there is a constant \ $\tilde{c}$ \ such that every set of  diameter \ $d$ \ can be covered by at most \ $\tilde{c}$ \ sets of diameter at most \ $d/2$. \ 
It may be of interest to also notice that a complete doubling space admits  a doubling measure [17]. \\

At a related level, one may wish to observe that quasisymmetric maps do not generally preserve the Hausdorff dimension of a set. Consider the metric 
space
\ $(X, d)$. \ Its snowflaked version, is by definition, the metric space \  $(X, d^\epsilon )$ \ where \ $\epsilon \in (0, 1)$. \ Using  H\"{o}lder's inequality we can 
see that 
the snowflaked metric \ $d^\epsilon$ \ obeys the triangle inequality, hence it is indeed a metric. Then it was proved [19] that each snowflaked version of
a doubling metric space can be  bi-Lipschitzly, hence \ $t^\epsilon$-quasisymmetrically embedded in some Euclidean space. Although this theorem 
characterizes the bi-Lipschitz   embeddings into Euclidean spaces it has a considerable drawback: roughly speaking, the curves on the snowflaked 
space are non-rectifiable [20]. Hence the snowflake map, although quite 
general for characterizing the embeddability of classes of quasi-symmetric maps, ``wrinkles" too much the underlying length structure of a metric space.      
This can also be seen, by noticing that if the Hausdorff dimension of \ $(X, d)$ \ is \ $n$, \ then the Hausdorff dimension of the snowflake \ 
$(X, d^\epsilon)$ \ is \ $ \frac{n}{\epsilon}$, \ where $\epsilon \  \in \ (0, 1)$. \ Hence the snowflake transformation can be used to produce fractals.
When we recall that the motivation for the functional form of the Tsallis entropy lies in multifractals
then we see that by constructing the quasisymmetric field isomorphisms \ $\tau_q$ \ we are being lead back, full circle, to the foundations of nonextensive 
Statistical Mechanics. In such considerations fractal configurations or self-similar and hierarchical structures figure prominently as, for instance, in the 
highly desirable dynamical justification of the form of the Tsallis entropy [2].\\
                                     
                                    \vspace{5mm}

%%%%%%%%%%%%%%%%%%%%%%%%%%%%%%%%%%%%%%%%%%%%%%%%%%%%%%%%%%%%%%%%%%%%%%%%%

      \centerline{\large\sc 4. \  Discussion and extensions}

                                    \vspace{3mm}

In this work we have constructed a generalized multiplication \ $\otimes_q$ \ which is distributive with respect to the generalized addition of the Tsallis 
entropy
$\oplus_q$. \ This allowed the construction of a deformed version of the basic sets used in elementary arithmetic. The deformation maps \ $\tau_q$ \ are 
intentionally 
constructed in such a way as to be a field isomorphisms, thus reproducing the structure of the undeformed sets, but with operations that reflect better the 
composition 
properties of the Tsallis entropy. We hope, that despite the seemingly arcane definition of \ $\otimes_q$, \  which turns out to be quite simple after thinking a 
bit a about it,
the field isomorphisms \ $\tau_q$ \ may be of some interest in elucidating the origin and in applications of the Tsallis entropy to particular systems.    
Unlike our previous work [7], here we recover in the expected manner the usual addition and multiplication in the limit of zero deformation \ $q\rightarrow 1$. 
\ As it  befits any mathematical structure, an obvious question is how general such a construction is or whether there is any more general or more natural way 
to  define these operations. This under the explicit proviso that greater generality should maintain or increase their potential use in applications of physical 
interest in nonextensive Statistical Mechanics.\\

The deformation maps \ $\tau_q$ \ seem to be quite intriguing from a metric and measure theoretical viewpoint. We have, in a very concrete way,
been driven around, by our constructions to the very concepts that motivated  the formulation of the Tsallis entropy in Physics; fractals and self-similar 
structures. Although we have presented some general comments that allow us to make the connection with fractals a bit more plausible, we have not really 
delved in the  present work into the heart of the matter as this would require an exploration of the underlying issues in Cartesian products 
of \ $\mathbb{R}_q$ \ and comparisons with the 
corresponding results of \ $\mathbb{R}^n$. \ Although in the present work we are interested in dealing with the ``one-dimensional" case \ $\mathbb{R}_q$, \ 
a vague outline of the underlying structures starts appearing.
A more complete understanding and appreciation of the significance of this construction requires the generalization of the concept of Cartesian products for 
\ $\mathbb{R}_q$ \ and the subsequent exploration  of the metric and measure theoretical of \ $\tau_q$ \ on such spaces. We feel that the underlying 
geometric and analytic 
framework should be along the spirit and techniques explored in [21], [22].  An issue of potentially greater impact for nonextensive Statistical Mechanics 
would be to investigate how the ergodic hypothesis formulated on $\mathbb{R}_q$-based dynamical systems can looks the viewpoint of structures based on
\  $\mathbb{R}$. \ This approach may provide a better understanding of the dynamical origin of the different values of \ $q$, \ but most concretely of the $q$-
sensitivity  
of the underlying dynamical  system to initial conditions. It may also contribute toward uncovering some of the conjectured relations between the different 
values of \ $q$ \ 
describing different aspects for a particular system. If such relations generically exist, they could be seen as a nonextensive Statistical Mechanics analogue 
of the relations among critical exponents of  systems undergoing a continuous phase transition in equilibrium Statistical Mechanics, which may help classify 
the systems  described by the Tsallis entropy into non-extensive ``universality classes". \\     

                                \vspace{7mm}

%%%%%%%%%%%%%%%%%%%%%%%%%%%%%%%%%%%%%%%%%%%%%%%%%%%%%%%%%%%%%%%%%%%%%%%
 
                                                    \centerline{\large\sc Acknowledgements}
   
                                                                          \vspace{3mm}
   
 \noindent After the completion of the present work, we became aware of the existence of  reference [3]  which provides a seemingly different  
 construction,  leading to the same main algebraic result, as the present work. We are grateful to Professor Constantino Tsallis 
 and to the anonymous referee for pointing out  reference \ [3] \  of which we were completely unaware during the course  of the present work. \\

                                                                         \vspace{7mm}
   
%%%%%%%%%%%%%%%%%%%%%%%%%%%%%%%%%%%%%%%%%%%%%%%%%%%%%%%%%%%%%%%%%%%%%%%%                             

      \centerline{\large\sc References}
      
                                \vspace{4mm}

\noindent [1] \ C. Tsallis, \ J. Stat. Phys. {\bf 52}, 479 (1988).\\
\noindent [2] \  C. Tsallis, \ Introduction to Nonextensive Statistical Mechanics: \\
                           \hspace*{6mm}  Approaching a Complex World, \ Springer (2009).\\
\noindent [3] \ T. C. Petit Lob\~{a}o,  P.G.S. Cardoso, S.T.R. Pinho, E.P. Borges, \ Braz. J. Phys. \\ 
                         \hspace*{6mm} {\bf 39}, 402 (2009).\\  
\noindent [4] \ O. Penrose, \ Foundations of Statistical Mechanics: A Deductive treatment, \\
                          \hspace*{6mm} Pergamon Press (1970).\\      
\noindent [5]  \  L. Nivanen, A. Le M\'{e}haut\'{e}, Q.A. Wang, \  Rep. Math. Phys. {\bf 52}, 437 (2003).\\
\noindent [6]  \  E.P. Borges, \  Physica A {\bf 340}, 95 (2004).\\
\noindent [7] \  N. Kalogeropoulos, \  Physica A {\bf 356}, 408 (2005).\\
\noindent [8] \ C. Tsallis, \  \emph{Some Open Points in Nonextensive Statistical Mechanics}, \  arXiv:1102.2408\\
\noindent [9] \ B. Lesche, \  J. Stat. Phys. {\bf 27}, 419 (1982).\\ 
\noindent [10] \ S. Abe, B. Lesche, J. Mund, \ J. Stat. Phys. {\bf 128}, 1189 (2007).\\ 
\noindent [11] \ S. Abe, \ Phys. Rev. E {\bf 66}, 046134 (2002).\\
\noindent [12] \ E.M.F. Curado, C. Tsallis, \ J. Phys. A {\bf 24}, L69 (1991);\\
                          \hspace*{8mm}  Ibid. {\bf 24}, 3187 (1991); \ \ Ibid. {\bf 25}, 1019 (1992).\\
\noindent [13] \ C. Tsallis, R.S. Mendes, A.R. Plastino, \ Physica A {\bf 261}, 534 (1998).\\ 
\noindent [14] \ M. Gromov, \ Milan J. Math. {\bf 61}, 9 (1991).\\
\noindent [15] \  A.C. Thompson, \  Minkowski Geometry, Cambridge Univ. Press (1996).\\  
\noindent [16] \ P. Tukia, J. V\"{a}is\"{a}l\"{a}, \ Ann. Acad. Sci. Fenn. Ser. A I, Math. {\bf 23}, 525 (1998).\\ 
\noindent [17] \ J. Heinonen, \  Lectures on Analysis on Metric Spaces, \ Springer-Verlag (2001).\\
\noindent [18] \ A. Beurling, L.V. Ahlfors, \ Acta Math. {\bf 96}, 125 (1956).\\
\noindent [19] \ P. Assouad, \ Bull. Soc. Math. France {\bf 111}, 429 (1983).\\
\noindent [20] \  J.T. Tyson, J.-M. Wu, \  Ann. Acad.Sci. Fenn. Ser. A I, Math. {\bf 30}, 313 (2005).\\ 
\noindent [21] \ G. David, S. Semmes, \ Fractured Fractals and Broken Dreams: \\ 
                          \hspace*{8mm} Self-Similar Geometry through Metric and Measure,  \ Clarendon Press (1997).\\
\noindent [22] \ S. Semmes, \ Some Novel Types of Fractal Geometry, Oxford Univ. Press (2001).\\

                          \vfill

\end{document}